\documentclass[%
 aip,
 jmp,%
 amsmath,amssymb,
 reprint,%
]{revtex4-1}

\usepackage{graphicx}
\usepackage{dcolumn}
\usepackage{bm}

\begin{document}


\title{Fabrication and Characterization of Controllable Grain Boundary Arrays in Solution Processed Small Molecule Organic Semiconductor Films}


\author{Songtao Wo}
\author{Randall~L. Headrick}
\affiliation{Department of Physics and Materials Science Program, University of Vermont, Burlington VT 05405}

\author{John E. Anthony}
\affiliation{ Department of Chemistry, University of Kentucky, Lexington, KY 40506}


\date{\today}

\begin{abstract}
We have produced solution-processed thin films of 6,13-bis(triisopropyl-silylethynyl)  pentacene with grain sizes from  a few micrometers up to millimeter scale by lateral crystallization from a rectangular stylus. Grains are oriented along the crystallization direction, and the grain size  transverse to the crystallization direction depends inversely on the writing speed, hence forming a regular array of oriented grain boundaries with controllable spacing. We utilize these controllable arrays to systematically study the role of large-angle grain boundaries in carrier transport and charge trapping in thin film transistors.  The effective mobility scales with the grain size, leading to an estimate of the potential drop at individual large-angle grain boundaries of more than one volt. This result indicates that the structure of grain boundaries is not molecularly abrupt, which may be a general feature of solution processed small molecule organic semiconductor thin films  where relatively high energy grain boundaries are typically formed. This may be due to the crystal Transient measurements after switching from positive to negative gate bias or between large and small negative gate bias reveal reversible charge trapping with time constants on the order of 10 s, and trap densities that are correlated with grain boundary density.   We suggest that charge diffusion along grain boundaries and other defects is the rate determining mechanism of the reversible trapping.
\end{abstract}



\maketitle

\section{\label{sec:introduction}Introduction}

Conjugated polymers and organic small molecules such as polythiophenenes\cite{Horowitz99,Katz01} and pentacenes\cite{Dimitrakopoulos96,Gundlach97, Fritz04} are of current interest because of their potential for low cost large area fabrication of various electronic devices such as solar cells,\cite{Walzer04,Ogane09,Lim09} light-emitting diodes\cite {Aernouts02} and thin film transistors.\cite{Sonar08,Singh05,Park07,Lin971,Lin972}  Organic semiconductor devices have advanced rapidly  in terms of overall performance since mobilities exceeding 1 cm$^{2}$V$^{-1}$s$^{-1}$ were first demonstrated in pentacene thin film transistors.\cite{Klauk02,Lin971}  In addition, an understanding of the fundamental mechanisms related to bias stress and other charge trapping effects is emerging.\cite{Lee10}  There has also been continuing progress in producing materials and processing methods with better overall uniformity and lifetime.\cite{Kang08,Ohe09}  Research on new methods for solution processing remains a key area, as well as understanding how such processes produce desirable thin film properties. 

Laterally directed film deposition from liquid solutions can produce highly oriented films. \cite{Tracz2003, Headrick2011, Headrick08, Becerril2008, DeLongchamp09} Drop casting with controllable drying can also produce films with a high degree of local orientation, but with circular symmetry on a mesoscopic scale.\cite{Chen08,Chen09} A number of other methods have also been devised which give some degree of orientation and control of grain structure.\cite{Rivnay09} Our method utilizes directed deposition from a hollow stylus composed of a rectangular glass capillary, and is based on a process that has been previously shown to produce oriented films with large grain size.\cite{Headrick08} We have recently developed the method further in order to vary the grain size through variation in the writing speed and substrate temperature. In this paper, we demonstrate the control of grain structure, the effect of grain boundaries on carrier transport, and reversible charge trapping for  organic field-effect transistors fabricated with 6,13-bis(triisopropyl-silylethynyl)(TIPS)  pentacene as the semiconductor material.
\begin{figure}[htbp]
   \centering
   \includegraphics[width=3.25in]{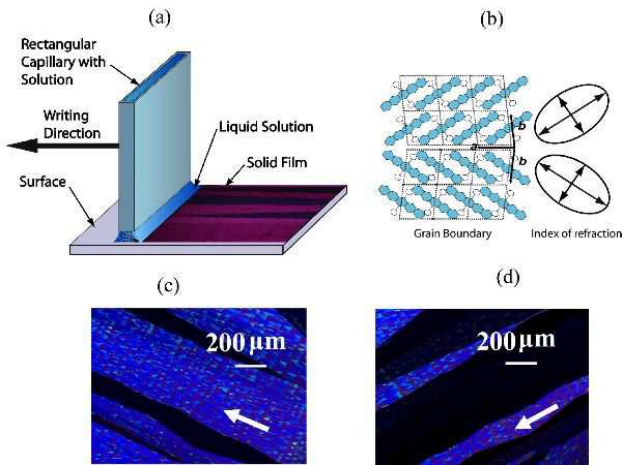}
   \caption{Schematic of hollow capillary writing process and the well defined twin boundaries. (a) The scheme of writing process; the writing direction is indicated by the black arrow.  (b) Molecular structure on the substrate,  looking down along the c* direction, the Fresnel (refractive index) ellipsoid is indicated by the oval, the short arrow is the low index axis, so called fast axis, and the long arrow is the high index axis (slow axis) which is align up  with the molecular direction. (c,d) A typical polarized image of thin film with large grain size, the contrast shows the grains with different orientation, grains are separated by the well defined twin boundaries.  The angle between the two images is 56$^\circ $. The thin film was made by 5 wt$\% $ in toluene with writing speed of 0.1 mm/s with thickness of 35 nm, white arrow indicating the writing direction. 
 }
   \label{fig:ExptLayout}
\end{figure}

In Sec. III we describe results that show that semi-periodic arrays of parallel grain boundaries can be obtained, typically with just two dominant grain orientations in a twin configuration leading to a well-defined grain boundary since the orientation of the grain on each side of the boundary as well as the orientation of the boundaries themselves are aligned with the writing direction. These structures make it possible to systematically characterize the effect of the grain boundaries on thin film transistor carrier mobility, which we discuss in Sec. IV. Small numbers, as few as four grain boundaries oriented so that they maximally impede the carrier flow are found to reduce the measured field effect mobility by a significant factor. This effect is shown to be highly reproducible over arrays of transistors with channels oriented parallel and perpendicular to the film writing direction, and the effect also depends on the orientation of the boundaries.

Studies of transient phenomena in organic thin film transistors have previously found that effects can be divided into reversible and irreversible mechanisms, i.e. mechanisms with time scales of $\sim $10 s vs. mechanisms with time scales that can reach many hours.\cite{Salleo2005} The slow mechanism is often referred to as the bias stress effect, and has been studied for a number of organic semiconductors such as polythiophene,\cite{Salleo2005} pentacene, \cite{Gu08} rubrene, tetracene, and TIPS-pentacene.\cite{Lee10} The bias stress mechanism has recently been described in terms of a model based on hole trapping through drift into the dielectric layer with very low mobility.\cite{Lee10} Here we assume this model, although we note that other models such as formation of bipolarons in the semiconductor have been proposed.\cite{Street03}

In Section \ref{sec:transienteffect} we investigate the reversible component. The effect has been linked to detrapping of electrons, which, when trapped induce positive mobile carriers in the transistor channel.\cite{Gu05,Gu08,Knipp09} Note the counterintuitive nature of this process, where electron trapping leads to an \emph{excess} of mobile positive carriers. The effect is convincingly dominated by electron  trapping (as opposed to hole trapping) because the accumulation of induced mobile holes due to the trapped electrons  induces a threshold shift to positive gate voltage. Gu $et$.$al$. have proposed that this phenomenon is related to water at the semiconductor/dielectric interface,\cite{Gu08} which is known to be present at dielect/organic semiconductor interfaces.\cite{Wo06} In Sec. V (A), we show that the electron detrapping effect is also observed in our films and $is$ $correlated$ $with$ $grain$ $boundary$ $density$. We thus propose that the rate limiting mechanism is diffusion of electrons into and out of the grain boundaries and other associated defects. Evidence for positive charge trapping is found in Sec. \ref{sec:transienteffectB} from reversible transient currents on switching between two different \emph{negative} gate voltages, where electron traps are assumed to be mainly empty in both conditions.\cite{Benor2008}

In Sec. VI, our measurements are discussed in terms of other work in the literature. These studies generally support our observation of reversible trapping of positive charge but they also clearly show that the charge does not necessarily only reside at grain boundaries.\cite{Jaquith09} This  is consistent with our results, since we also observe measurable trap density for films with no large-angle grain boundaries (although there can be small angle grain boundaries or other defects such as dislocations), but at a significantly lower level than the films with large-angle grain boundaries. Shallow trap states at grain boundaries appear to play a role in producing a current bottleneck, thus reducing the effective mobility.

\section{\label{sec:exptmethods}Experimental methods}

TIPS-pentacene is synthesized as previously reported.\cite{Anthony01,Anthony02} The deposition technique used in this study is a direct-write method using a hollow rectangular capillary with size of 0.5 $ \times $ 5.0 mm$^2$ I.D. (Wale apparatus Co.$\#$4905-100) as depicted in Fig. 1 (a), A solution with a concentration of 0.2 - 5.0 wt$\% $ in toluene is held in the capillary by capillary forces. The capillary is bent at 1 cm from the end to form an "L" shape so that a gravity feed at constant pressure corresponding to the height of the vertical part is maintained.  The  capillary can hold enough solution in the horizontal part to coat a one inch long substrate, eliminating the need for solution pumping. Filling and cleaning of the capillary is straightforward since  the narrow cross section causes it to draw in solution by capillary action.  A substrate is mounted on a computer-controlled linear translation stage (Newport, M-VP-25XA) equipped with a thermoelectric module for temperature control.  Film deposition is accomplished by allowing the microdroplet of solution on the end of the capillary to make contact with the surface and then laterally translating the substrate at a controlled rate, typically 0.02-25 mm/s. 

A heavily doped n-type Si substrate with a 300 nm thermally grown SiO$_2$ layer is degreased in acetone and methanol, and rinsed with hot acetone and hot isopropyl. To make thin film transistors, the semiconductor layer is deposited on top of the SiO$_2$ with thickness of 30 - 50 nm. Gold is evaporated through a shadow mask to form source and drain electrodes with a 100 $\mu$m channel length and 900 $\mu$m channel width, creating a "bottom gate, top-contact" transistor geometry in which the  substrate serves as the gate contact.

\begin{figure}[htbp]
   \centering
   \includegraphics[width=3.0in]{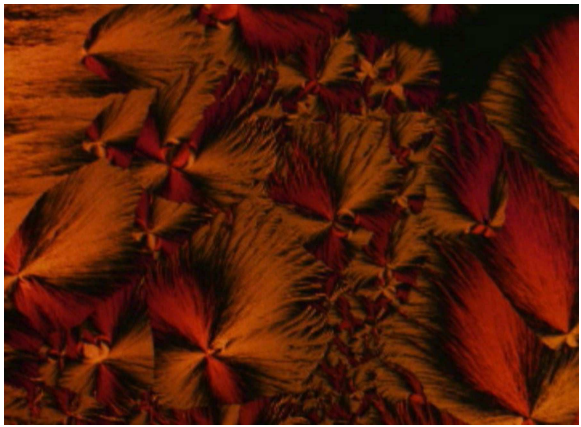}
   \caption{Frame from a video microscope movie taken during film deposition at 4 mm/s. (enhanced online) The image size is 0.53 mm $\times$ 0.40 mm. 
 }
   \label{fig:growthmovies}
\end{figure}

Polarized optical microscopy (Olympus, BXFM) is used to produce images of the thin films. We can identify the fast axis (low refractive index axis) of the grain using a full wave plate (gypsum, $\Delta$= 530 nm) thereby determining the grain orientation. Movies of the growth process can also be produced with the same microscope, as shown in Fig. \ref{fig:growthmovies}.

The electrical measurements are carried out on a  microprobe stage (Cascade M150) with Keithley 2636 dual source meter units in the dark with N$_2$ protection. The field-effect mobility is measured at both saturation and linear regime by transfer characteristic measurement. For the saturation regime, the gate bias is swept from $-$60 V to 60 V and back to $-$60 V when drain-source voltage  fixed at $-$60 V. The drain  current is:
\begin{equation}\
\
I_{d,sat}  = \frac{W}{2L}\mu_{sat} C_i (V_g  - V_t)^2,
\
\end{equation}
where $I_{d,sat}$, $W$, $L$, $V_g$, $V_t$, $C_i$ are the drain current, channel width, channel length, gate voltage, threshold voltage and capacitance per unit area (here we use $C_i$=10.0 $ nF/cm^{ - 2} $).  So $\mu_{sat}$ can be extracted from the plot $I_{d,sat}^{1/2}$ vs.$V_g$.  In the linear regime, the $V_d$ is fixed at $-$10V and drain current is:
\begin{equation}\
\
I_{d,lin}  = \frac{W}{L}\mu_{lin} C_i (V_g  - V_t )V_d,
\label{eqn:lineardraincurrent}
\end{equation}
so $\mu_{lin}$ can be extracted from  $I_{d,lin}$ vs.$V_g$.

Charge trapping/detrapping is studied by transient effect measurements which as mentioned above, are done in two ways: (i) Drain source current is measured under $V_g$  biased at 60 V for 300 s and abruptly switched to $-$20 V with fixed  $V_d=-$20 V. (ii) $V_g$ was biased at $-$50 V for 300 s and switched to $-$20 V when $V_d$=$-$20 V. The trap density can be estimated from the transient current based on the model that we discuss  in Sec.V.

\begin{figure}[h]
   \centering
   \includegraphics[width=1.5in]{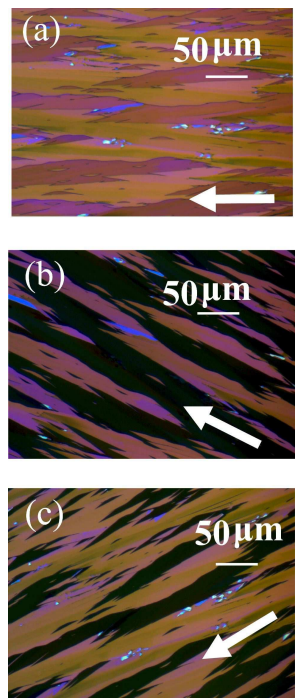}
   \caption{Grain structure of a 25 nm thick  film made from a 5 wt$\% $ solution at 0.2 mm/s. (a) Thin film is placed at 0$^\circ $ with the writing direction (b) Thin film rotated by 28$^\circ $ clockwise to show one type of grain when its reflection is extinct. (c) Thin film rotated by 28$^\circ $ counter-clockwise to show the twin  grains when in extinction.  The average grain size  perpendicular to the writing direction is 20 $\mu $m.
 }
   \label{fig:grainstructure}
\end{figure}

\section{\label{sec:grainstructure}Grain structure and twin boundary}
Fig. 1 (b) is the schematic of the well defined twin boundary.  When the thin film is deposited, the [001] direction is perpendicular to the substrate.  Fig. 1 (b) shows the top view of the molecular structure. Note that there are  grains with two different orientations. They are separated by a twin boundary which runs  parallel to the writing direction. The grain orientation is such that the direction of highest mobility is closely aligned along the writing direction. The Fresnel ellipsoid is depicted, illustrating that the molecular long axis is aligned  with the optical slow axis.  Assuming the bulk crystal structure, \cite{Anthony01} the angle between the molecular long axis and [100] is 28$^\circ $. So by using polarized microscopy, the reflected light intensity is expected to become extinct upon rotating the sample by $ \pm 28^\circ $ relative to the writing direction. The direction of the rotation depends on whether the grains are in the twinned orientation or not. Fig. 1(c, d) shows that  thin films made from 5 wt$\% $ solution at a speed of 0.1 mm/s with large grain size are in fact rotated by $ \pm 28^\circ $. The extinct grains of the two images are  complementary indicating a well defined growth orientation along the [100]  crystallographic axis with alternating twin grain orientations. The average grain size perpendicular to the writing direction is $L_g \approx $ 200 $\mu $m.

When the writing speed is increased,  grain nucleation occurs more frequently and the grain size is reduced laterally but the growth orientation and twin grain boundary orientations stay the same. Fig. \ref{fig:grainstructure} shows a thin film made at a coating speed of 0.2 mm/s. As shown in Fig. \ref{fig:grainstructure} (b) and (c), the extinction in the two images are exactly complementary.\\

\begin{figure}[h]
   \centering
   \includegraphics[width=3in]{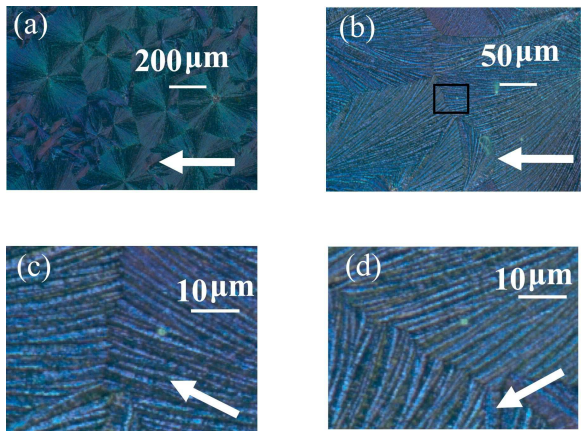}
   \caption{Grain structure of a 35 nm thick film made from a  5 wt$\% $ solution at 2.0 mm/s. (a,b) Thin film is placed at 0$^\circ $  with the writing direction with 5$ \times $ and 20$ \times $ objective lenses (c) The zoom in image of the region inside the black rectangle of (b), with the sample rotated 28$^\circ $ clockwise. There is a particle serving as a mark, the grain it sitting on extinct, the other two next to it is bright (d) The  zoom in image of the black box of (b) by rotating sample 28$^\circ $ counter-clockwise. The grain under the particle reverses the contrast as well the other two next to it which imply the same type grain boundary between the "stripe"  and average width of the stripe is 3$ \pm$1 $\mu $m.
 }
   \label{fig:fanlikestructure}
\end{figure}

When the writing  speed is further increased above a critical value, the dynamic meniscus become much larger as the solution is pulled out of the capillary by viscous forces leaving a wet film; then the crystallization no longer occurs at a well defined contact line.\cite{Wo2011}  Instead, nucleation occurs randomly to form a spherulitic grain structure as shown in Fig. \ref{fig:fanlikestructure}.  A video of the crystallization process is also included in Fig. \ref{fig:growthmovies} (online only). Images (c) and (d) are from the  region within the black rectangle in  image (b), but rotated by $ \pm 28^\circ $; a particle near the center of this region serves as a mark.  When the sample is rotated  28$^\circ $ clockwise, the grain under the particle is dark while the two grains next to it are bright. When it is rotated 28$^\circ $ counterclockwise, the contrast reverses completely showing that the grain orientation of each "stripe" is also along  the [100] and the grain boundary is the same twin grain boundary observed at the lower speeds. 

\section{\label{sec:effectofgb}Effect of grain boundaries on mobility}
\subsection{\label{sec:mobilityanisotropy}Electrical measurement and mobility anisotropy}
\begin{figure}[h]
   \centering
   \includegraphics[width=3in]{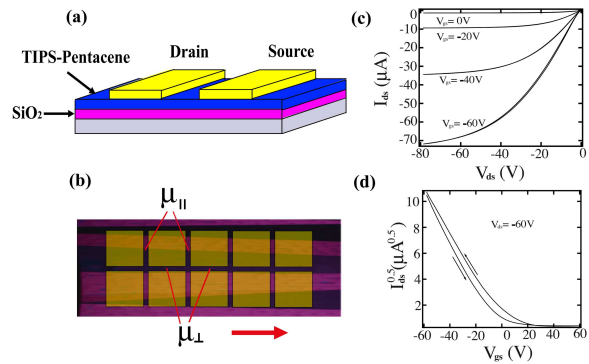}
   \caption{(a) Typical top contact device configuration used, 300nm thick thermal oxide is served as insulator. Gold electrodes are thermally evaporated on top of the active layer with a shadow mask as drain and source contacts with channel length  $L$=100 $\mu $m and width $W$=900 $\mu$m. (b) Schematic of an array of quad structure used to measure the mobility of the transistor both in direction parallel and perpendicular to the writing direction. The actual structures consist of a 20$\times$2 array.  (c,d) I-V output/transfer characteristic of a sample with mobility 0.8 cm$^{2}$V$^{-1}$s$^{-1}$. }
   \label{fig:deviceconfiguration}
\end{figure}

A fourty gold pad "quad" structure is utilized for the anisotropy measurement depicted in Fig. \ref{fig:deviceconfiguration}. Each film drawn from  the  TIPS-pentacene organic solution  covers an array of 58 devices including 38 of them in parallel with the writing direction and 20  in the perpendicular direction.    This structure allows us to measure the mobility on both orthogonal directions rapidly over arrays of transistors so that a statistical analysis can be carried out. The I-V output and transfer characteristics of a device with mobility 0.8 cm$^{2}$V$^{-1}$s$^{-1}$ are depicted in  Fig. \ref{fig:deviceconfiguration} (c) and (d) respectively. Fig. \ref{fig:deviceconfiguration} (d) shows a positive threshold and noticeable hysteresis implying a large number of  trap states which we will discuss  in detail.\\
 \begin{figure}[h]
   \centering
   \includegraphics[width=3in]{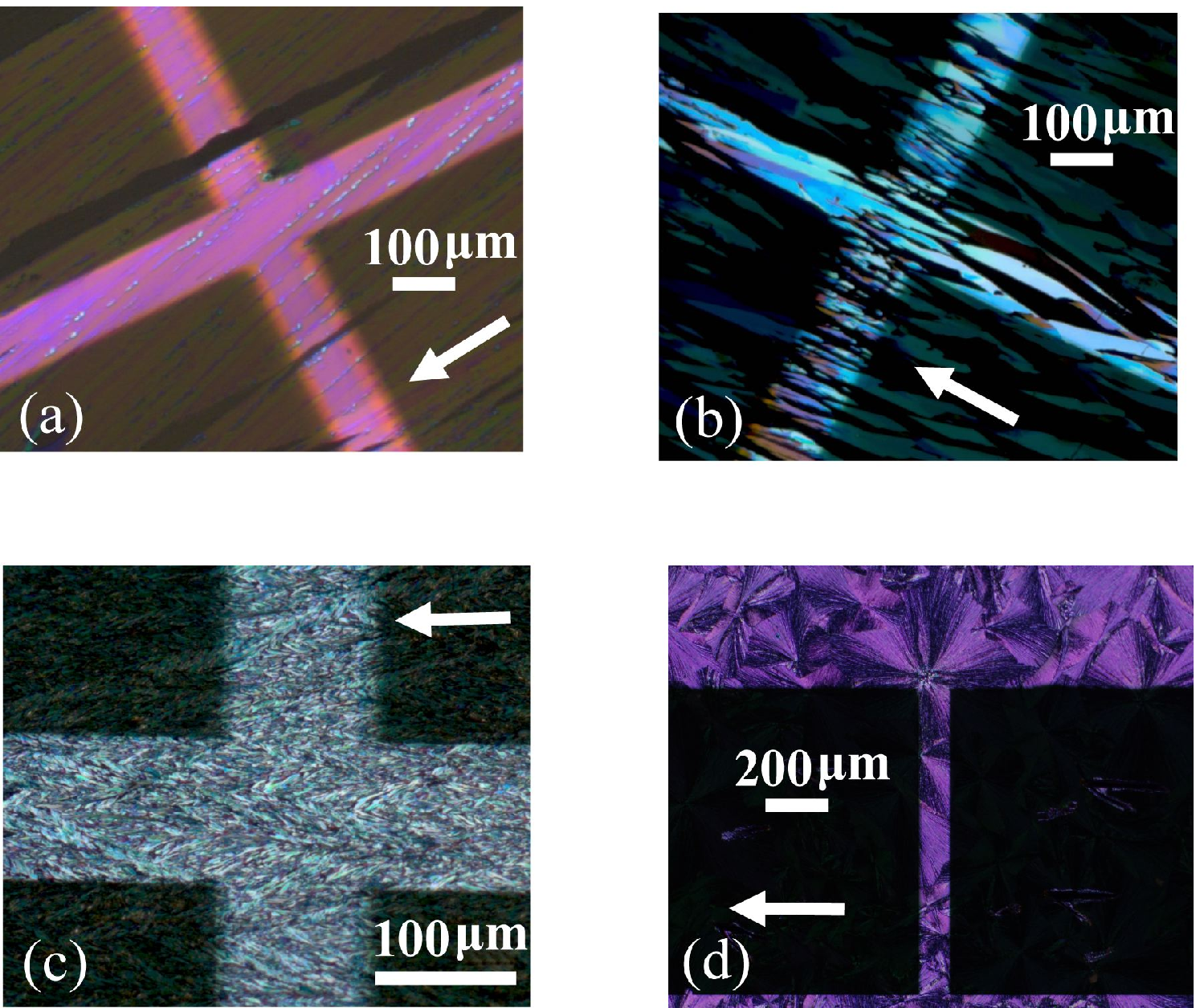}
   \caption{Transistors with different grain structure by varying the speed with 5 wt$\% $ solution. (a) 0.06 mm/s, $\mu _\parallel  $
=0.29 cm$^{2}$V$^{-1}$s$^{-1}$, $\mu _ \bot  $=0.17 cm$^{2}$V$^{-1}$s$^{-1}$. $L_g  > 200 \mu$m. (b) 0.2 mm/s, $\mu _\parallel  $=0.29 cm$^{2}$V$^{-1}$s$^{-1}$, $\mu _ \bot  $=0.07 cm$^{2}$V$^{-1}$s$^{-1}$.  $L_g  =$20$ \pm $4 $\mu $m. (c) 1.0 mm/s, $\mu _\parallel  $=0.058 cm$^{2}$V$^{-1}$s$^{-1}$, $\mu _ \bot  $=0.037 cm$^{2}$V$^{-1}$s$^{-1}$.  $L_g  =$8$ \pm $2 $\mu $m. (d) 2.0 mm/s, $\mu _\parallel  $=0.015 cm$^{2}$V$^{-1}$s$^{-1}$, $\mu _ \perp  $=0.015 cm$^{2}$V$^{-1}$s$^{-1}$.  $L_g  = 3 \pm 1 \mu $m. The mobility is strongly correlated with the grain structure in terms of the grain size.}
   \label{fig:transistorswithdifferentgrainstructure}
\end{figure}

Fig. \ref{fig:transistorswithdifferentgrainstructure} shows that by varying the substrate speed,  thin film transistors with different well controlled grain structures can be made. The transistors are made of 5 wt$\% $ solution with speed of (a) 0.06 mm/s, (b) 0.2 mm/s, (c) 1.0 mm/s, (d) 2.0 mm/s respectively. The difference of the grain structure is striking. At low speed, because of the well defined contact line of the meniscus, large grains are formed as shown in Fig. \ref{fig:transistorswithdifferentgrainstructure} (a).\cite{Wo2011} When the writing speed  is increased, grain nucleation occurs more frequently and the grain size is reduced especially in the direction perpendicular to the writing direction as shown in Fig. \ref{fig:transistorswithdifferentgrainstructure} (b). Because the grains are long enough to bridge to the channel in the parallel direction the mobility remains at a high value, but the mobility in the direction perpendicular to the writing direction is reduced by a factor of 2.4. When the writing speed is increased above 1.0 mm/s, as shown in Fig. \ref{fig:transistorswithdifferentgrainstructure} (c) the grain size is further reduced in both directions, and the mobility is correspondingly reduced in both directions. At 2.0 mm/s, the spherulitic grain structure shown in Fig. \ref{fig:transistorswithdifferentgrainstructure} (d) exhibits the very special "stripe" structure described above. The mobility is the same in both direction because the orientation of the grains in the channel is nearly random. Comparing the grain structure and mobilities at 0.06 mm/s and 0.2 mm/s (Fig. \ref{fig:transistorswithdifferentgrainstructure} (a) and (b)), $\mu _\parallel$ is almost unchanged, while $\mu _\perp$ is reduced by a factor more than 2 due to the presence of  four grain boundaries across the channel, on average.  This effect illustrates convincingly that the grain boundaries are a bottleneck for carrier transport. In order to estimate the potential drop on each grain boundary we adopt a series resistance model with the transistor working in the linear regime:
\begin{equation}\
\
\\V_{d}  =\Delta  V_{grain}  + N\,\Delta V_{GB},
\
\end{equation}
where $\Delta V_{grain} $, $\Delta V_{GB}$, $N$ are the potential drop on all the grains and potential drop on each grain boundary. Eq. (3) can also be rewritten in terms of effective mobility $\mu _{eff}$, intragrain mobility $\mu _{0}$, and grain boundary resistance $R_{GB} $. Using $q_e\,n_0 = C_i V_g$ inserted into Eq.(2) for $I_{d,lin}$, where $n_0$ is the mobile carrier density per unit area induced by gate  and $\Delta V_{GB}  = I_{d,lin} R_{GB}$,  we predict a linear relationship between $1/\mu _{eff} $ and the number of grain  boundaries $N$:
\begin{equation}\
\
\frac{1}{\mu _{eff} } = \frac{1}{\mu _0 }(1 + Nf),
\label{eqn:mastergbequation}
\end{equation}
where $f = \mu _0 R_{GB} \,q_e\,n_0W/L$. Thus, it is expected that the mobility will scale  with the grain size. To test this idea, linear mobility is measured with $V_d=-$10 V for the four samples of Fig. \ref{fig:transistorswithdifferentgrainstructure}, and the estimated grain boundary number $N$ and the linear mobility of the perpendicularly oriented channel are listed in Table I. It confirms that the linear mobility scales inversely with the number of grain boundaries blocking the channel.  A value $f$ = 0.33$ \pm $0.05 is estimated from the data in Table \ref{tab:linearmobility}. For the case of four grain boundaries, $N$ = 4, given $\mu _0$ = 0.1, $\mu _{eff}$ = 0.04 and $V_d = -$10 V, we can estimate that $R_{GB}\approx 10^6  \Omega$ and $\Delta V_{GB} \approx$ 1.3 V. This large voltage drop on the grain boundary cannot occur within a 1 nm distance because it would lead to an electric field on the order of $10^9$ V/m, i.e.  larger than the breakdown field for most materials. This surprising result suggests that the effective grain boundary thickness is on the scale of 10 nm or more. We note that Teague $et$.$al$. found a potential drop across a grain boundary of about 1V by surface potential imaging.\cite{Teague08} Therefore, such a finding is not unprecedented.  However it is significant that we conclude that this is the \emph{typical behavior} of grain boundaries in this materials system when deposited from solution.  Given that there are a great many studies of carrier transport in vapor deposited films that show high mobility even for very small grain size, even at the nanometer scale,\cite{Lin971} we speculate that there is a fundamental difference between grain boundaries in solution processed films vs. vapor deposited films, i.e. grain boundaries in small molecule solution processed films appear to present a larger barrier to carrier transport.  

\begin{table}[htbp]
   \centering
\begin{tabular}{c|cccccc}
\hline\hline
sample& speed& $N$ &$\mu _{ \bot ,linear}$&$f$~&$\mu _{ \parallel ,sat}$&$\mu _{ \bot ,sat}$\\
 & (mm/s)&  &($cm^2 /V$-$s)$&&($cm^2 /V$-$s)$&($cm^2 /V$-$s)$\\
 \hline
a& 0.06~~~~& 0~~ &0.1~~~~&$-$~&0.29&0.17\\
b& 0.2~~~~& 4~~ &0.04~~~~&0.38~&0.29&0.07\\
c& 1.0~~~~& 12~~ &0.02~~~~&0.33~&0.058&0.037\\
d& 2.0~~~~& 33~~ &0.01~~~~&0.28~&0.015&0.015\\
\hline\hline
\end{tabular}
\caption{Mobility measurement for individual transistors with different grain structures, as  depicted in Fig. \ref{fig:transistorswithdifferentgrainstructure}.  Here $N$ is the average number of grain boundaries oriented to block the perpendicular channel, estimated from polarized optical images. }
   \label{tab:linearmobility}
   \end{table}

\begin{figure}[h]
   \centering
   \includegraphics[width=2in]{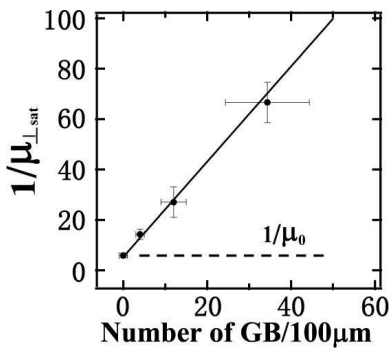}
   \caption{From the estimation of the grain size $L_g$  and the saturation mobility measured in the perpendicular direction, we find a linear relationship between 1/$\mu _ \bot  $  and number of grain boundaries. The dashed line indicates the level of  1/$\mu _0$. }
   \label{fig:MobvsGb}
\end{figure}

In the saturation regime Eq.(4) should still be valid aside from the numerical value of $f$.  Therefore, we also examine the correlation between the saturation mobility  and the grain size.   Fig. \ref{fig:MobvsGb} shows a clear linear relationship between $1/\mu _{eff} $ and grain boundary number $N$, again implying the validity of the series resistance  model. 

\subsection{\label{sec:transistorarrays}Transistor array measurements}
\begin{figure}[h]
   \centering
   \includegraphics[width=3in]{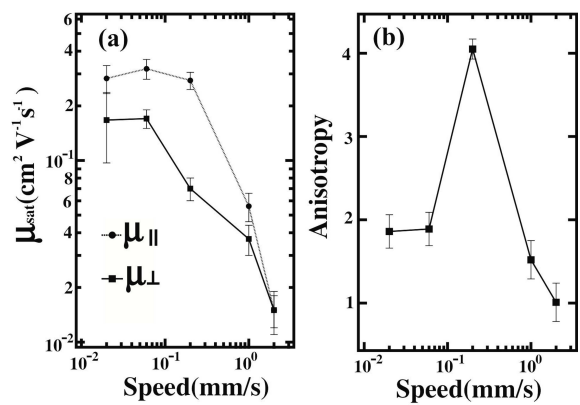}
   \caption{The statistical collection of mobility measurements as a function of writing speed. (a) Each data points on the plot is an average value from 20 (38) measurement for $\mu _{ \bot ,sat}$ ($\mu _{ \parallel ,sat}$).  Small error bar indicates good repeatability and  uniformity of the film.  The anisotropy value extracted from (a) is depicted in (b) We found an intrinsic anisotropy ratio  of 1.8. The anisotropy is enhanced at 0.2 mm/s when there are four grain boundaries across the perpendicular transistor channel but essentially none in the parallel channel.}
   \label{fig:Mobvsspeed}
\end{figure}

Saturation mobility of the parallel and perpendicular channel on arrays  of devices  is shown in Fig. \ref{fig:Mobvsspeed}. Each data point is a statistical collection of 38 mobility measurements for $\mu _{ \parallel ,sat}$ and 20 for $\mu _{ \bot ,sat}$.  Note that at 0.02 mm/s a relatively large error bar for the mobility is found because the grain orientation has a significant variation.  Since a single grain dominates each transistor, the angle variation is translated to a variation in the measured mobility due to the intrinsic mobility anisotropy of the material.  All other data exhibit small variations, showing that the  thin film fabrication method is highly repeatable. Thus,  the difference of the mobility as the speed is varied is predominantly due to the change of the grain structure. The intrinsic mobility anisotropy $\mu _\parallel$/$\mu _ \bot  $ is 1.8 which can be obtained from the measurement at 0.06 mm/s when the grain boundary effect is insignificant.   This value is in good agreement with the photoconductivity measurement by Ostroverkhova.\cite{Ostroverkhova06} Interestingly, when the writing speed is 0.2 mm/s and there are four grain boundaries across the channel perpendicular to the writing direction, a pronounced reduction of the mobility on $\mu _ \bot  $ is found while   $\mu _\parallel$ remains same. The anisotropy value is greatly enhanced to 4 as depicted at Fig. \ref{fig:Mobvsspeed}(b). \emph{This result clearly shows that the effect of as few as four grain boundaries cutting the channel produce a significant bottleneck, reducing the effective mobility by a factor of $\approx $ 2.5.} When the speed is increased, the grain size is reduced to less than the channel length in both parallel and perpendicular direction  so  the mobility is reduced in both directions. At  a writing speed of 2.0 mm/s, the mobility becomes isotropic and the grains exhibits a spherulitic structure (which can be quite striking visually, as seen in Fig. \ref{fig:growthmovies},  \ref{fig:fanlikestructure} and Fig. \ref{fig:transistorswithdifferentgrainstructure} (d)). The low value of the mobility is due to the very small grain size.\\

\section{\label{sec:transienteffect}Transient effect measurement}
In subsection \ref{sec:transienteffectA}, the transient effect of the devices are measured while switching from the  off state ($V_g  = +60$ V for 300 s, to ensure that the electron traps are maximally filled) to the on state ($V_g  =  - 20$ V). We observe two different effect: (i) ``slow" bias stress effect, with time constant $ \sim 10^4 $ s. Here we adapt the model recently proposed by Lee \emph{et. al}, that is, the decay of the current is caused by holes drifting into the insulator.\cite{Lee10} (ii) ``fast'' reversible electron trapping and detrapping, which is believed to be mainly responsible for the hysteresis in transfer curves.\cite{Gu05,Gu08} This effect is correlated with the grain boundary density with time constant $ \approx 40 $ s.

In subsection \ref{sec:transienteffectB}, the transient effect of the devices are characterized while switching from  $V_g  =  0$ to $- 50$ V and then from  $V_g  = -50$ V to $  - 20$ V with $V_{ds}  = -20$ V.  This measurement is designed to be sensitive to hole traps, since when switching between two different negative gate voltages, the electron traps should be mainly depopulated in both states.\cite{ucurum2008}  A reversible effect is observed with time constant $ \approx10 $ s, possibly due to holes diffusing into traps at the grain boundaries and other defects. 

\begin{table*}[htbp]
\centering

\begin{tabular}{cccccccccccc}
\hline\hline
Sample &orient.& Temp &speed& conc. & $\mu _{sat} $~~& $\mu _{lin} $~~ & $\tau _{dr} $~~~& $\beta _{dr} $~~& $n _{hi}(0)$~~& $\tau_{et} $~~&$\beta_{et} $~~~~\\

 &~&($^\circ$C)&(mm/s)&(wt$\% $)&$({\rm{cm}}^{\rm{2}} {\rm{V}}^{{\rm{ - 1}}} {\rm{s}}^{{\rm{ - 1}}} )$
  &$({\rm{cm}}^{\rm{2}} {\rm{V}}^{{\rm{ - 1}}} {\rm{s}}^{{\rm{ - 1}}} )$&(s)~~&~~
  &$({\rm{10}}^{{\rm{11}}} {\rm{cm}}^{{\rm{ - 2}}} )$
  &~(s)~~ &~~ \\
\hline
A&$\parallel$ &25&2& 5& 0.02 & 0.02&  8000~~~~& 0.51~~~~& 3.30   & 40~~ & 0.50~~~~ \\

B&$\parallel$ &25&2& 5& 0.015 & 0.013&  8000~~~~ & 0.51~~~~& 3.55  & 40~~   & 0.50~~~~ \\

D&$\parallel$&25 &2& 5& 0.013& 0.011 &  20000~~~~  & 0.55~~~~& 4.68   & 30~~   & 0.55~~~~ \\

~E$^*$&$\perp$&25  &2& 5& 0.0048& 0.0024&  15000~~~~ & 0.50~~~~& 10.27   & 40~~   & 0.54~~~~ \\

C&$\parallel$ &25&0.2& 5& 0.28 & 0.129&  35000~~~~ & 0.55~~~~& 1.24   & 15~~   & 0.55~~~~ \\

~C$^*$&$\perp$&25 &0.2& 5& 0.07& 0.029 &  8000~~~~& 0.54~~~~& 2.24   & 50~~   & 0.50~~~~ \\



H&$\parallel$&25 &0.08& 5& 0.31& 0.127 &  26000~~~~& 0.55~~~~& 1.19  & 22~~   & 0.53~~~~ \\




N&$\parallel$&60 &0.1& 0.3& 0.80 & 0.19&  8000~~~~& 0.51~~~~& 1.50   & 22~~   & 0.55~~~~ \\




~L$^*$&$\perp$&60  &0.1& 0.3& 0.13 & 0.048&  35000~~~~& 0.55~~~~& 1.77   & 28~~   & 0.55~~~~ \\
\hline\hline

\end{tabular}
\caption{Fitting parameter for the transient current when $V_g$ was switched from $+60$ V to $-20$ V with $V _{d}=-20$ V. Heated samples were heated to 60$^\circ $ during TIPS-pentacene deposition. Note the abbreviated subscripts:  $sat =$ saturation, $lin =$ linear, $dr =$ drift, $hi =$ induced holes, $et =$ electron trap.}\label{tab:bigtableoffittingparameters}
   \end{table*}

\subsection{\label{sec:transienteffectA}Bias stress effect and electron trapping at grain boundaries}

Transient phenomena were studied  to understand the relationship between the trap states and grain structure. First we describe a simple model that we use to separate the slow bias stress from the reversible effects. The key feature of the model is that the density of  mobile carriers $n _{mob}(t)$ change with time when the device is turned on.  In general, this can be either due to electron traps discharging or hole traps charging.    The density of charged electron traps as a function of time is represented as  $n _{et}(t) $, and these trapped electrons induce an equal density of mobile holes  $n_{hi}(t) $ = $n _{et}(t) $ thereby adding to the drain current.  Conversely, charged hole traps represented as $n_{ht}(t) $ remove carriers from participating in current flow.  Combining these terms with the total gate charge density $n_0$, which depends only on the gate voltage and capacitance, we have:

 \begin{equation}
n _{mob}(t)  = n _0  - n _{ht}(t) +n _{hi}(t).
\
\end{equation}

In order to model the slow bias stress effect, we follow Lee's prescription that $d\,n_{0}/dt \propto  - n_{0}^2$ if the effect is dominated by drifting holes driven  into the dielectric by the electric field.   This model leads to a stretched hyperbola form,  after considering that the trap states in the dielectric follow an exponential distribution of energy levels.  For the reversible effect, we assume that  the transient effect is dominated by the discharging of electron traps, so that $d\,n_{et}/dt \propto  - n_{et}$. This equation leads to a stretched exponential form, again with the assumption of an exponential tail of trap states.  In this case the electron traps are assumed to be somewhere in the semiconductor layer, which will be in the process of discharging after switching the device from the off state to on. Finally we have, in the approximation that the time scales of the drift and electron detrapping processes are very different:

\begin{equation}
\
n_{mob}  \approx \frac{{n_0 }}{{1 + (t/\tau _{dr} )^{\beta _{dr} } }} + n_{hi} (0)\exp ( - (t/\tau_{et} )^{\beta_{et} } ).
\label{eqn:transeqnA}
\end{equation}
  
\begin{figure}[htbp]
   \centering
   \includegraphics[width=3in]{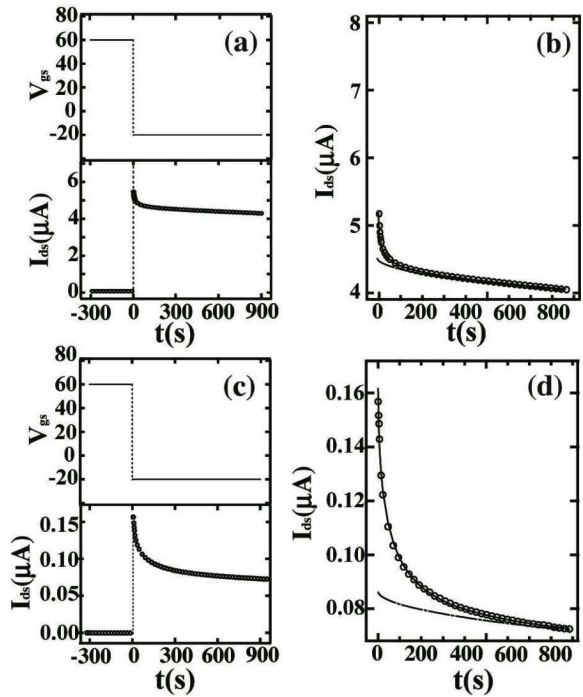}
   \caption{Transient effect measurement with switching gate bias from 60 V to $-$20 V. (a) Large grain size ($L$$ > $200 $\mu $m) with mobility of 0.28 cm$^{2}$V$^{-1}$s$^{-1}$. (b) The later 900s measurement of (a) when gate bias is $-20$ V after bias with 60 V for 300 s, long dashed below the data points is a single stretched hyperbola function with $\tau _{dr} $ = 35000 s and $\beta _{dr} $ = 0.55,  solid line is the sum of the stretched hyperbola function and stretched exponential function. (c) Small grain size ($L = 1 \sim 2 \mu $m) with mobility 0.0048 cm$^{2}$V$^{-1}$s$^{-1}$. (d) The later 900 s measurement of (b) when gate bias is $-$20 V after bias with 60 V for 300 s, long dashed line below the data points is a single stretched hyperbola function with $\tau _{dr} $ = 15000 s and  $\beta _{dr} $ = 0.55, solid line is the sum of the stretched hyperbola function and stretched exponential function. }
   \label{fig:timescanelectrondetrapping}
\end{figure}

By fitting the transient current curve with Eq. (\ref{eqn:transeqnA}), the trap density can be extracted and is listed in Table \ref{tab:bigtableoffittingparameters} for representative samples.  In order to model the drain current, the expression for $n_{mob}(t)$ above is inserted  into  Eq. (\ref{eqn:lineardraincurrent}) with $ C_i V_g \rightarrow q_e\,n_{mob}(t)$.  Eq. (\ref{eqn:transeqnA}) also requires the linear mobility, which was measured from transfer  curves with $V_{ds}  = -6$ V for each sample.  Table \ref{tab:bigtableoffittingparameters} lists  extracted parameters for selected samples, including the stretching exponents, which were found to be in the typical range for similar processes.\cite{Lee10}    The  extracted values of $n _{hi}(0)$  show  devices with smaller mobility values have larger trap state densities, suggesting a correlation with the presence of grain boundaries. 

In Fig. \ref{fig:timescanelectrondetrapping} we present detailed results for samples C and E*.  Fig. \ref{fig:timescanelectrondetrapping} (a) and (b) show how the current decays after the gate bias is switched from 60 V (off state) to $-$20 V (on state) for sample C. When the gate is biased at 60 V, the electron trap states are filled and thus induce mobile holes. Once the gate bias is switched to $-$20V,  the trapped electrons depopulate, resulting in a decay of the induced hole density and hence of the drain current, with a time constant of 15 s (see Table \ref{tab:bigtableoffittingparameters}). The lower curve in (b) shows the curve corresponding to only the bias stress component (time constant 35000 s).  The full fit, including both terms of Eq. (\ref{eqn:transeqnA}) is seen to match the data almost perfectly, thus demonstrating the utility of the model in separating the two transient components. A relatively small effect of the reversible component is observed, corresponding to an electron trap density, when fully filled, of $\approx 10^{11}$ cm$^{-2}$.  We consider sample C, which has one of the lowest trap densities of all of the samples that we measured, to be characteristic of samples with a nearly single crystal semiconductor layer.  Fig. \ref{fig:timescanelectrondetrapping} (c) and (d)  show results for the extreme opposite case, sample E*, which has a very small grain size and a grain morphology similar to the one shown in Fig. \ref{fig:transistorswithdifferentgrainstructure} (d). We extract an electron trap density of $\approx 10^{12}$ cm$^{-2}$. In this case,  we interpret the time constant $\tau_{et}$  as being characteristic of electron detrapping.  If we assume that most of the extra traps are associated with the presence of grain boundaries, then the density of traps per grain boundary, with $N \approx 30$ for sample E*, is on the order of $10^{10}$ cm$^{-2}$. This is a remarkably large density in the sense that, taking into account geometric factors such as the area of the transistor channel and the thickness of the semiconductor film (35 nm), the density of trap states on each grain boundary is calculated to be  $> 10^{13}$ cm$^{-2}$.  Based on electrostatic considerations, this density is probably too high to be present at a planar boundary, However, since traps are spatially localized on grain boundaries rather than being randomly distributed over the channel, they can form a bottleneck at a lower overall average density.  Therefore, the calculation above may be an overestimate. This possibility is discussed below in Sec. \ref{sec:discussion}.

Table \ref{tab:bigtableoffittingparameters} also lists results for  heated substrates.  Due to the increased evaporation rate of the solvent at elevated temperatures we can produce large grain size and high mobility at higher speed when the substrate is heated.   Surprisingly the carrier mobility  is more than double the highest value that we have measured previously. However, the trap density deduced from transient measurement is not reduced compared to lower mobility large grain samples prepared at room temperature.  This indicates that the difference in mobility is not correlated with  a reduction in trap states. Note that we are comparing large grain samples with minimal grain boundaries at both temperatures, so that the difference is not explained by an apparent difference in grain structure.

\subsection{\label{sec:transienteffectB}Reversible hole trapping/detrapping}

\begin{table*}[htbp]

   \centering

\begin{tabular}{ccccccccccc}
\hline\hline
sample & speed& concentration  & $\mu _{sat} $~~& $\mu _{lin} $~~&$n _{0}-n _{ht }(\infty)$~~~~& $\tau _{dr} $~~~~& $\beta _{dr} $~~~& $n _{ht}(\infty) $~~~~~& $\tau _{diff} $~~~~&$\beta _{diff} $~~~~\\

 &(mm/s)&(wt$\% $)&$({\rm{cm}}^{\rm{2}} {\rm{V}}^{{\rm{ - 1}}} {\rm{s}}^{{\rm{ - 1}}} )$&$({\rm{cm}}^{\rm{2}} {\rm{V}}^{{\rm{ - 1}}} {\rm{s}}^{{\rm{ - 1}}} )$
  &(${\rm{10}}^{{\rm{11}}} {\rm{cm}}^{{\rm{ - 2}}}) $&(s)~~~~&~~~~
  &$({\rm{10}}^{{\rm{11}}} {\rm{cm}}^{{\rm{ - 2}}} )$
  &(s)~~~~  &~~~~ \\
\hline
C &0.2& 5& 0.28 & 0.129& 32.1& 5500~~~~~& 0.60~~~~& NA  & NA~~~~~   & NA~~~~~ \\

E$^* $  &2& 5& 0.0048& 0.0024&30.8& 7000~~~~~& 0.60~~~~& 2.30   & 13~~~~~& 0.55~~~~~ \\
\hline\hline
\multicolumn{10}{l}{${}^*$devices are measured at perpendicular channel.}\\
\end{tabular}
\caption{Fitting parameter for the transient current when $V_g$ switch from $-$50 V to $-$20 V with $V _{d}=-$20 V.}
   \label{tab:morefittingparameters}
   \end{table*}

\begin{figure}[h]
   \centering
   \includegraphics[width=3in]{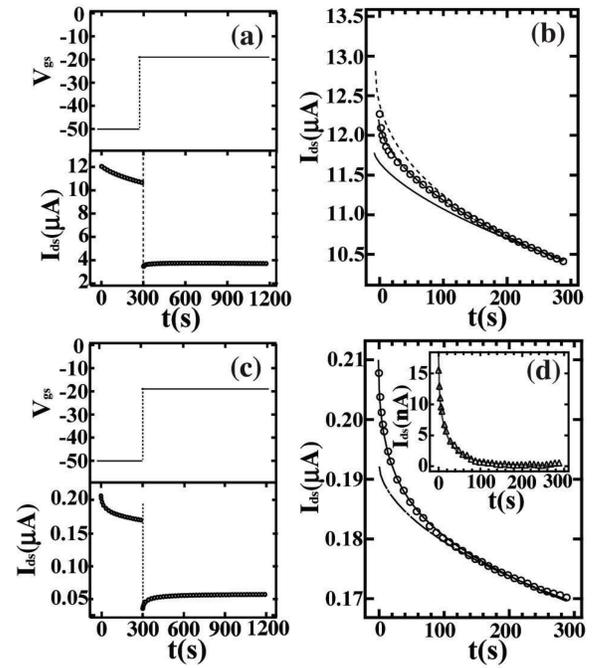}
   \caption{Transient measurement with switching gate bias from $-$50 V to $-$20 V. (a) Large grain size($L$$ > $200 $\mu $m) with mobility
            of 0.28 cm$^{2}$V$^{-1}$s$^{-1}$. (b) The first 300 s measurement of (a) when gate bias is -50 V, long dashed line
            crossing the data points is a single stretched hyperbola function with $\tau _{dr} $=5500 s and$\beta _{dr} $=0.6, the
            dashed line is for  $\beta _{dr}$=0.5 and the solid line is for  $\beta _{dr}$=0.7. (c) Small grain size($L = 1 \sim 2 \mu
            $m) with mobility 0.0048 cm$^{2}$V$^{-1}$s$^{-1}$. (d) The first 300 s measurement of (b) when gate bias is $-$50 V,
            solid line is the sum of the stretched hyperbola function and stretched exponential function. The long dashed line is
            the stretched hyperbola function only. The inset shows the stretched exponential component of the fit, along with the data minus 	the bias stress component.
   }
   \label{fig:transientholetrapping}
\end{figure}

Another variation of the transient effect measurement is carried out with switching between two negative gate voltages.  In Fig.  \ref{fig:transientholetrapping} data for  samples C and E* is presented. Compared to Fig. \ref{fig:timescanelectrondetrapping} this measurement brings the Fermi level closer to the valence band maximum, where hole traps are more likely to be found.   The drain current is measured starting when the gate bias is switched from 0 V to  $-$50 V(on state).  After 300 s the gate is switched to $-$20 V  for 900 s.  Fig. \ref{fig:transientholetrapping} (a) and (b) show the results for sample C.   A gradual decrease in drain current is observed when the device is first turned on.  After switching to the smaller gate voltage, a small increase is observed over the first $\approx$ 60 s, followed by stable operation.  Fig. \ref{fig:transientholetrapping} (c) and (d) show the same measurement for sample E*. The data shows a much larger relative change over the first minute compared to sample C.  It is clear from inspection of (d) that  the trapping/detrapping process is  reversible with the same time constant in each direction. Furthermore, the similar magnitude of the effect in each direction suggests that most of the trap states exist at energies relative to the valence band maximum that are accessible between the two negative gate voltages.  Therefore, it is plausible that the transient effect observed at negative gate voltages correspond mainly to filling and emptying of hole trap states.   
 
We use the same model developed in Sec. \ref{sec:transienteffectA} to fit this data.  A minor difference is the assumption that hole traps are being filled, as discussed above.  Therefore, after switching to $-$50 V ($t = 0$ in the figure)  $ n_{ht}$ ($ht =$ holes trapped) is extracted from the fitting rather than $n_{hi}$ ($hi =$ holes induced). Taking the hole trap density to be  $n_{ht}(\infty)$  and assuming that  $\tau _{dr} \gg \tau _{diff}$,

\begin{equation}
\
n_{mob}  \approx \frac{{n_0  - n_{ht} (\infty )}}{{1 + (t/\tau _{dr} )^{\beta _{dr} } }} + n_{ht} (\infty )\exp ( - (t/\tau _{diff} )^{\beta _{diff} } )).
\
\end{equation}\\

Note that this expression is for the time interval with $V_d = -$50 V.  We have not attempted to fit the later time part because the state of the interface with regard to the bias stress effect is not well known when switching from a higher field to a lower one.   The data in Fig.  \ref{fig:transientholetrapping} (b)  is consistent with a single stretched hyperbola function, i.e.  due to the drifting of the mobile holes into the insulator layer. The time constant $\tau _{dr}  =$  5500 s, is shorter than the time constant found for  $V_g =  -20$ V (Table \ref{tab:bigtableoffittingparameters}).  This difference is consistent with the effects being due to drift, since the time constants are expected to be shorter when the field strength is higher.\cite{Lee10}  The complete set of parameters extracted from the model are listed in Table \ref{tab:morefittingparameters}.

In order to fit the data in Fig.  \ref{fig:transientholetrapping} (d),  a second component corresponding to the reversible process is deduced.  The time constant is 13 s,  which we suggest to be characteristic of hole diffusion into grain boundaries. Note that the recovery after switching the gate to $-20$ V is also consistent with this picture since detrapping of holes with a nearly identical time constant neatly explains the observed increase in the drain current.  A similar model based on hole detrapping  has been proposed to explain transient measurements of vapor deposited pentacene field-effect transistors.\cite{ucurum2008}
 
\subsection{\label{sec:transienteffectC}Correlation of trap states with grain boundaries}
\begin{figure}[h]
   \centering
   \includegraphics[width=3.4in]{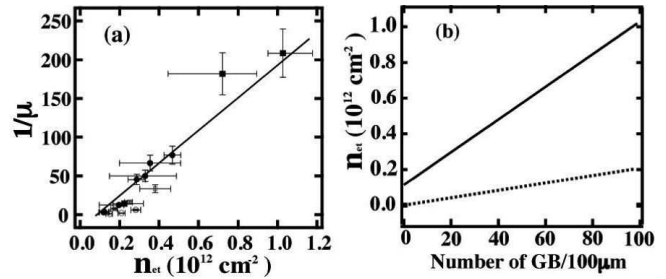}
   \caption{(a) The electron trap density extracted from the transient effect measurement with multiple samples: parallel channel without heating the substrate (filled circle); perpendicular channel without heating the substrate (filled square); parallel channel with heating the substrate (open circle); perpendicular channel with heating the substrate (open square). The solid line is derived from Fig. \ref{fig:MobvsGb} given a simple linear relation between the $n_{et}$ and number of grain boundaries. (b) The linear correlation $n _{et}  = 1.0 \times 10^{10} N + 1.2 \times 10^{11}$ , where $N$ is the number of grain boundaries. The dotted line corresponds to a similar estimate for hole traps.}
   \label{fig:gbvsmobandtraps}
\end{figure}

Fig. \ref{fig:gbvsmobandtraps} (a) shows the electron trap density ($n _{et} $) extracted from transient current data as described above, and  plotted versus the inverse of the saturation mobility. A linear relationship is observed, suggesting that the traps are  correlated with the mobility.  This is not believed to be a causal relationship, and instead   both are believed to be directly determined by the grain boundary density.  In Fig. \ref{fig:MobvsGb}, we found the linear relationship between $1/\mu {}_{eff}$ and the number of grain boundaries in the channel $N$. Combining the results of Fig. \ref{fig:gbvsmobandtraps} (a) with  Fig. \ref{fig:MobvsGb},  a simple linear relation between  $n_{et}$ and number of grain boundaries is deduced as shown in the  Fig. \ref{fig:gbvsmobandtraps} (b).  Note that since this is a combination of results from two data sets, we omit the data points and simply show a plot of the correlation. The trap density per grain boundary  is estimated to be $1.0 \times 10^{10}$ cm$^{ - 2}$, and the intercept value, $1.2 \times 10^{11}$ cm$^{ - 2}$,  is the electron trap density due to the defects not associated with grain boundaries. The dashed line in Fig. \ref{fig:gbvsmobandtraps} (b) is the same plot for hole traps.  In this case, the density is $2 \times 10^{9}$ cm$^{ - 2}$\ per grain boundary.  Since the traps are thought to be inhomogeneously distributed in the channel because they are associated with grain boundaries, a more natural unit is per grain boundary length. Multiplying by the channel length (100 $\mu$m), the hole trap line density is $\lambda_h \approx$ 2 nm$^{-1}$ along a grain boundary.  However, this latter value probably overestimates the local trap density as we discuss below.\\

\section{\label{sec:discussion}Discussion} 
 
 In this section we discuss the results in terms of possible models for the grain boundary structure.  First, we note that the large grain boundary resistance deduced in Sec. \ref{sec:mobilityanisotropy} is consistent with conducting probe atomic force microscopy results on vapor deposited sexithiothene singe grain boundaries.\cite{Kelley01}  Several previous studies have noted a correlation between grain size and mobility,\cite{Kang04,Chen08} or have deduced the effect of grain boundaries by varying the transistor channel length in organic field effect transistors.\cite{Gundlach08}  Such effects have been particularly prevalent, and problematic, in solution deposited films.   Therefore, we suggest that high grain boundary resistance is a typical feature of solution processed organic semiconductor films, although this may be more closely related to the crystal packing than to the processing conditions $per se$, as we explain below.  Conversely, there are many examples of vapor deposited films where grains size appears to be of secondary importance,\cite{Shtein02} pointing to the possibility that vapor deposited films are less susceptible to such effects.

It has been suggested by Rivnay et al that high grain boundary resistance can be related to crystal packing since herringbone-type molecular films can readily have low molecular misorientation across large-angle boundaries.\cite{Rivnay09} On the other hand $\pi - $stacked crystal structures, of which TIPS-Pentacene is an example,  are thought to be most susceptible to having only large molecular misorientation across grain boundaries, leading to high grain boundary resistance. We can extend this idea by pointing out that many vapor-deposited organic semiconducors have a herringbone arrangement, which accounts for the general trends alluded to above.

Our data is relevant to this discussion because in order to hold the large voltage drop that we observe, the boundary cannot be molecularly abrupt as generally assumed.  Following the crystal packing arguments of Rivnay et al, we assume that the dominant grain boundary that we depict in Fig. \ref{fig:ExptLayout} (b) is a relatively high energy structure because of the large (56$^{\circ}$) misorientation of the molecules across the interface.  In this case a high degree of disorder near the boundary follows logically from these arguments, since the molecules near the boundary may become rotated or displaced in order to reach a more favorable local configuration. While there may well be other effects that influence the grain boundary resistance, e.g. related to the processing conditions, this simple model appears to be consistent with our results, as well as a large body of the existing literature.
 
 Hole traps at grain boundaries have been reported by several groups in which spatial maps of trapped positive charge has been observed.\cite{Puntambekar06,Tello08,Jaquith09} Our observation of hole traps correlated with grain boundaries is consistent with these studies.  However, our estimates of trap densities appear  to overestimate the trap densities after accounting for the localization of the traps in the region of the grain boundaries.   Since the grain boundaries are a bottleneck to current flow, the drain current should be very sensitive to trap densities that alter the mobile carrier concentration in the grain boundary. But, given that the transient effects that we observe in Figs. $\ref{fig:timescanelectrondetrapping}$ and $\ref{fig:transientholetrapping}$ only modulate the drain current by a small amount, the density of trapped charge producing the transient effect must be small compared to total density of mobile carriers, i.e. $<  10^{12}$ cm$^{-2}$ \emph{in the grain boundary region}.  If we estimate the width of the grain boundary to be 10 nm, then the line density is $\lambda_h < 0.1$ nm$^{-1}$, which is considerably lower than the value estimated above.
  
Note that this effect can also be recast in terms of a barrier model since it produces  a local reduction in mobile carriers by electrostatic repulsion from fixed charge, which is the essential mechanism of the grain boundary barrier model.\cite{Horowitz00,Nussbaumer98,Seto75,Orton80,Street02}  However, we reiterate  that since the transient effects are weak, barriers cannot be the primary cause of the large grain boundary resistance. Therefore, a simple trap model without significant barrier effects is the most natural way to explain the drastically reduced mobility when grain boundaries are present.\cite{Street02}  This is plausible because a high density of shallow traps is likely to accompany the relatively deep traps that are responsible for the reversible transient currents. This model can be implemented by assuming two different mobilities, one for the grains ($\mu_0$) and the other for the disordered region within the grain boundaries ($\mu_{GB}$).   With this model, we can extend the discussion of Sec. \ref{sec:mobilityanisotropy}, and find a complementary form for the $f$ factor in Eq. (\ref{eqn:mastergbequation}), from which we can estimate the grain boundary mobility,
  
  \begin{equation}\
\
f = \frac{L_{GB}}{L} \left( \frac{\mu_0}{\mu_{GB}} - 1 \right).
\
\end{equation}\label{eqn:mynewffactor}

A similar expression has been derived by Chen et al.\cite{Chen08}    We take the  grain boundary thickness $L_{GB}$ to be about 10 nm.  With $f = 0.33$ and $\mu_0 = 0.1$ cm$^2$ V$^{-1}$s$^{-1}$, we estimate the mobility within the grain boundaries to be $\mu_{GB} = 3 \times 10^{-5}$ cm$^2$ V$^{-1}$s$^{-1}$. This value is in line with typical values obtained for disordered organic semiconductor materials.\\

\section{\label{sec:conclusion}Conclusions}
Solution-processed TIPS-pentacene thin films are fabricated by a rectangular stylus.  The grain size of the thin films can be varied over a wide range,  from a few micrometers up to millimeter depending on the substrate speed and temperature, with  the [100] crystallographic axis of the grains aligned with the writing direction for speeds $\le$1 mm/s, and randomly oriented for higher speeds. Measurements of anisotropic mobility reveal that grain boundaries are a bottleneck for carrier transport. A potential drop at individual grain boundaries of more than a volt is deduced. Transient measurements reveal a reversible charge trapping associated with grain boundaries. Straightforward analysis of the data suggests that shallow traps within disordered grain boundaries with a grain boundary thickness  of at least 10 nm can explain the results, although transient currents are evidence for modest barrier effects.\\

\section{\label{sec:acknowledgement}Acknowledgement}
This material is based upon work supported by the National Science Foundation under Grant No.DMR-0722451 and DMR-0348354.


%

\end{document}